\documentclass{ptapap}

\author{Swayamtrupta Panda}[CFT,CAMK]
\author{Bo{\.z}ena Czerny}[CFT,CAMK]
\author{Conor Wildy}[CFT]
\author{Marzena {\'S}niegowska}[CFT,UW]
\affil[CAMK]{Nicolaus Copernicus Astronomical Center, ul. Bartycka 18, 00--716 Warsaw, Poland}
\affil[CFT]{Center for Theoretical Physics, Al. Lotników 32/46, 02--668 Warsaw, Poland}
\affil[UW]{Warsaw University Observatory, Al. Ujazdowskie 4, 88-478 Warsaw, Poland}

\title{Testing the physical driver of Eigenvector 1 in Quasar Main Sequence}
\begin{document}

\maketitle

\begin{abstract}

Quasars are among the most luminous sources characterized by their broad band spectra ranging from radio through optical to X-ray band, with numerous emission and absorption features. Using the Principal Component Analysis (PCA), \citet{bg92} were able to show significant correlations between the measured parameters. Among the significant correlations projected, the leading component, related to Eigenvector 1 (EV1) was dominated by the anti-correlation between the Fe${\mathrm{II}}$ optical emission and [OIII] line where the EV1 alone contained 30\% of the total variance. This introduced a way to define a quasar main sequence, in close analogy to the stellar main sequence in the Hertzsprung-Russel (HR) diagram (\citealt{sul01}). Which of the basic theoretically motivated parameters of an active nucleus (Eddington ratio, black hole mass, accretion rate, spin, and viewing angle) is the main driver behind the EV1 yet remains to be answered. We currently limit ourselves to the optical waveband, and concentrate on theoretical modelling the Fe${\mathrm{II}}$ to H$\mathrm{\beta}$ ratio, and test the hypothesis that the physical driver of EV1 is the maximum of the accretion disk temperature, reflected in the shape of the spectral energy distribution (SED). We performed computations of the H$\mathrm{\beta}$ and optical Fe${\mathrm{II}}$ for a broad range of SED peak position using CLOUDY photoionisation code. We assumed that both H$\mathrm{\beta}$ and Fe${\mathrm{II}}$ emission come from the Broad Line Region represented as a constant density cloud in a plane-parallel geometry. We compare the results for two different approaches: (1) considering a fixed bolometric luminosity for the SED; (2) considering $\mathrm{L_{bol}/L_{Edd}}$ = 1.
\end{abstract}

\section{Introduction}

Quasars are among the most luminous objects in the observable universe associated with rapidly accreting supermassive black holes. We classify AGN depending on how the AGN is viewed – whether the detected radiation allows us to „see” the nucleus directly (Type-1 AGNs); or the radiation is partially obscured due to the presence of the dusty torus lying in-between the observer and the source (Type-2 AGNs). In Type-1 AGNs, the continuum emission is dominated by the energy output in the optical-UV band that  comes from the accretion disk which surrounds a supermassive black hole \cite[e.g.][]{c87,cap2015} and the broad emission lines in the optical-UV, including Fe${\mathrm{II}}$ pseudo-continuum and Balmer Component are usually considered to be coming from the Broad Line Region (BLR) clouds. The spectral properties of the broad band spectra and the line emissivities are strongly correlated (\citealt{bg92,sul00,sul02,sul07,yip04,sh14,sun15}). The Principal Component Analysis (PCA) is a powerful tool that allows to procure the dominating correlations that can be used to identify the quasar main sequence. This sequence is analogous to the stellar main sequence in the Hertzsprung-Rusell diagram. Quasar main sequence was suggested to be driven mostly by the Eddington ratio (\citealt{bg92,sul00,sh14}), but also by the additional effect of the black hole mass, viewing angle and the intrinsic absorption (\citealt{sh14,sul00,kura09}). Among the significant correlations projected, the leading component, related to Eigenvector 1 (EV1) is dominated by the anti-correlation between the Fe${\mathrm{II}}$ optical emission and [OIII] line where the EV1 alone contained 30\% of the total variance. The parameter R$_{\mathrm{FeII}}$, which strongly correlates to the EV1, is the ${\mathrm{FeII}}$ strength, defined to be the ratio of the equivalent width of ${\mathrm{FeII}}$ to the equivalent width of ${\mathrm{H\beta}}$.
\par 
We postulate that the true driver behind the R$_{\mathrm{FeII}}$ is the maximum of the temperature in a multicolor accretion disk which is also the basic parameter determining the broad band shape of the quasar continuum emission. The hypothesis seems natural because the spectral shape determines both broad band spectral indices as well as emission line ratios, and has already been suggested by \cite{b07}. We expect an increase in the maximum of the disk temperature as the R$_{\mathrm{FeII}}$ increases. According to Figure 1 from \cite{sh14},  increase in R$_{\mathrm{FeII}}$ implies increase in the Eddington ratio or decrease in the mass of the black hole. We expect that this maximum temperature depends not only on the Eddington ratio \citep{c06}, but on the ratio of the Eddington ratio to the black hole mass (or, equivalently, on the ratio of the accretion rate to square of the black hole mass).

\section{Basic Theory}


The spectral energy distribution (SED) for a typical quasar reveals that most of the quasar radiation comes from the accretion disk and forms the Big Blue Bump (BBB) in the optical-UV (\citealt{c87, rich06}), and this thermal emission is accompanied by an X-ray emission coming from a hot optically thin mostly compact plasma, frequently refered to as a corona (\citealt{c87,haa91,Fabian2015}). The ionizing continuum emission thus consists of two physically different spectral components. We parametrize the BBB component by the maximum of the disk temperature, which according to the standard Shakura-Sunyaev accretion disk model, is related to the black hole mass and the accretion rate 

\begin{equation}
\mathrm{T}_{\mathrm{BBB}} = \left[\frac{3\mathrm{GM}\dot{\mathrm{M}}}{8\pi \sigma \mathrm{r}^3}\left(1 - \sqrt{\frac{\mathrm{R}_{\mathrm{in}}}{\mathrm{r}}}\right)\right]^{0.25} = 1.732\times 10^{19} \left(\frac{\dot{\mathrm{M}}}{\mathrm{M}^{2}}\right)^{0.25}, \label{eq:01}
\end{equation}
where $\mathrm{T}_{\mathrm{BBB}}$ - maximum temperature corresponding to the Big Blue Bump; G - gravitational constant; M - black hole mass; $\dot{\mathrm{M}}$ - black hole accretion rate; r - radial distance from the centre; $\mathrm{R_{in}}$ - radius corresponding to the innermost stable circular orbit. $\mathrm{M}$ and $\dot{\mathrm{M}}$ are in cgs units. Similar formalism has been used by \cite{b07} although the coefficient differs by a factor of 2.6 from Eq. (1). This maximum is achieved not at the innermost stable orbit around a non-rotating black hole (3R$_{\mathrm{Schw}}$) but at  4.08$\bar{3}$ R$_{\mathrm{Schw}}$.  The SED component peaks at the frequency

\begin{equation}
\nu_{\mathrm{max}} \sim \left[\frac{\frac{\mathrm{L}}{{\mathrm{L}_{\mathrm{Edd}}}}}{\mathrm{M}}\right]^{\mathrm{0.25}}\label{eq:02}.
\end{equation}
where $\nu_{\mathrm{max}}$ - frequency corresponding to $\mathrm{T}_{\mathrm{BBB}}$; L - accretion luminosity $\left (=\eta \dot{\mathrm{M}} \mathrm{c}^2 \right )$; $\mathrm{L_{\mathrm{Edd}}}$ - Eddington limit $\left (= \mathrm{\frac{4\pi GMm_{p}c}{\sigma_{T}}}\right )$, where $\mathrm{m_{p}}$ - mass of a proton, $\sigma_{\mathrm{T}}$ - Thompson cross section.
\par 
We use a power law with a fixed slope ($\alpha_{uv}$) for the accretion disk spectrum with an exponential cutoff which is determined by the value of $\mathrm{T}_{\mathrm{BBB}}$. The X-ray coronal component shape is defined by the slope ($\mathrm{\alpha_{x}}$) and has an X-ray cut-off at 100 keV (\citealt{fra02} and references therein). The relative contribution is determined by fixing the broad band spectral index $\mathrm{\alpha_{ox}}$, and finally the absolute normalization of the incident spectrum is set assuming the source bolometric luminosity. We fix most of the parameters, and $\mathrm{T}_{\mathrm{BBB}}$ is then the basic parameter of our model.
\par  
Some of this radiation is reprocessed in the BLR which produces the emission lines. In order to calculate the emissivity, we need to assume the mean hydrogen density ($\mathrm{n_H}$) of the cloud, and a limiting column density (N$_\mathrm{H}$) to define the outer edge of the cloud, and we use here a single cloud approximation. Ionization state of the clouds depends also on the distance of the BLR from the nucleus.  We fix it using the observational relation by \cite{b13}
\begin{equation}
\left(\frac{\mathrm{R}_{\mathrm{BLR}}}{1\;\mathrm{lt-day}}\right) = 10^{\left[1.555 + 0.542\; \mathrm{log}\left(\frac{\lambda \mathrm{L}_{\lambda}}{10^{44} \;\mathrm{erg s}^{-1}}\right)\right]}\label{eq:03}
\end{equation}
The values for the constants considered in Equation 3 are taken from the Clean $\mathrm{H \beta\; R_{BLR} - L}$ model from \citet{b13} where $\lambda$ = 5100 \AA.

\section{Results and Discussions}
In \cite{panda17}, we checked the dependence of the change in the R$_{\mathrm{FeII}}$ as a function of the maximum of the accretion disk temperature, T$_\mathrm{{BBB}}$  at constant values of L$_\mathrm{bol}$, $\mathrm{\alpha_{uv}}$ , $\mathrm{\alpha_{ox}}$ , $\mathrm{n_H}$ and $\mathrm{N_H}$. Here, we refer to the approach in \cite{panda17} as Method-1 (M1). 
\par
In M1, the source bolometric luminosity is fixed, $\mathrm{L_{bol}}$ = $\mathrm{10^{45} \;erg\; s^{-1}}$ with accretion efficiency $\epsilon$ = 1/12, since we consider a non-rotating black hole in Newtonian approximation (see Eq.~1). This determines the accretion rate, $\mathrm{\dot M}$. For a range of disk temperature from $5\times 10^4\;$K to $5\times 10^5\;$K, we computed the range of black hole mass to be between [$2.2\times 10^6\; \mathrm{M_{\odot}}$, $2.2\times 10^8\; \mathrm{M_{\odot}}$] using the Eq. (1). From the incident continuum, we estimated the $\mathrm{L_{5100\AA}}$ which we then used as an input to derive the $\mathrm{R_{BLR}}$ using the Eq. (3). The ratio of the optical to X-ray spectral index, $\mathrm{\alpha_{ox}}$ was fixed at -1.6 which specifies the optical-UV and X-ray luminosities. This allowed us to determine the normalization of the X-ray bump. The resulting two-power law SED was constructed with an optical-UV slope , $\mathrm{\alpha_{uv}}$ = -0.36, and X-ray slope, $\mathrm{\alpha_{x}}$ = -0.91 \citep{roz14} and their corresponding exponential cutoffs (the optical-UV cutoff is determined from the $\mathrm{T_{BBB}}$ while for the X-ray case it was fixed at 100 keV). We tested two cases by changing the mean hydrogen density from (i) $\mathrm{n_H = 10^{10}\; cm^{-3}}$ to (ii) $\mathrm{n_H = 10^{11} \;cm^{-3}}$, keeping the hydrogen column density, $\mathrm{N_H = 10^{24} \;cm^{-2}}$, in accordance to \cite{bv08}. We had dropped the X-ray power-law component in M1 in further computations for simplicity. Knowing the irradiation, we computed the intensities of the broad Fe${\mathrm{II}}$ emission lines using the corresponding levels of transitions present in CLOUDY 13.05 \citep{f13}. We calculated the Fe${\mathrm{II}}$ strength (R$_{\mathrm{FeII}}$ = EW$_{\mathrm{Fe{II}}}$ / EW$_{\mathrm{H\beta}}$), which is the ratio of Fe${\mathrm{II}}$ EW within 4434-4684 \AA $\;$to broad H$\beta$ EW. This prescription is taken from \cite{sh14}. All the simulations have been considered without including microturbulence.
\par 
In Method-2 (M2), we now make three important modifications: (i) we allow for the presence of the hard X-ray power law (ii) we fix the Eddington ratio (i.e., $\mathrm{L_{bol}/L_{Edd} = 1}$) instead of a constant bolometric luminosity (iii) we use the observational relation between the UV and X-rays to obtain $\alpha_{\mathrm{ox}}$. Fixing the Eddington ratio provides us with a relation between the mass accretion rate ($\mathrm{\dot{M}}$) and black hole mass ($\mathrm{M_{BH}}$). For the same range of maximum of the disk temperature as in M1, we calculate the black hole masses using the Eq.(1) which we have in the range [$6.06\times 10^5\; \mathrm{M_{\odot}}$, $6.06\times 10^9\; \mathrm{M_{\odot}}$].  We then determine the normalisation factor by integrating the optical-UV spectrum to the bolometric luminosity (which we calculate for each case from the range of $\mathrm{M_{BH}}$ as mentioned above). We then compute the values of $\mathrm{L_{2500\AA}}$ and $\mathrm{L_{5100\AA}}$ from the incident continuum. We use the $\mathrm{L_{2500\AA}}$ as the $\mathrm{L_{UV}}$ in the Eq. (1) from \cite{lusso17} to compute the $\mathrm{L_{X}}$ at 2 keV:
\begin{equation}
  \left\{\begin{array}{r@{}l@{\qquad}l}
(\log{\mathrm{L_X}}-25) = (0.610\pm 0.019)(\log{\mathrm{L_{UV}}}-25)+ \\
(0.538\pm 0.072)[\log{\mathrm{v_{FWHM}}}-(3 + \log{2})] + (-1.978\pm 0.100)
\end{array}\right.
\end{equation}
Here, $\mathrm{v_{FWHM}}$ is estimated using the corresponding $\mathrm{M_{BH}}$, $\mathrm{R}_{\mathrm{BLR}}$ and assuming a virial factor, f=1. Subsequently, the value of $\mathrm{\alpha_{ox}}$ is determined, i.e., $\mathrm{\alpha_{ox}} = -0.384\left(\frac{\mathrm{L_{2500}}}{\mathrm{L_{2keV}}}\right)$.  
\par 
The results from the photoionization modeling for both the methods (M1 and M2) are shown in Fig.1.

\begin{figure}[h!]
\begin{center}
\includegraphics[height=13cm, width=8cm, angle=270]{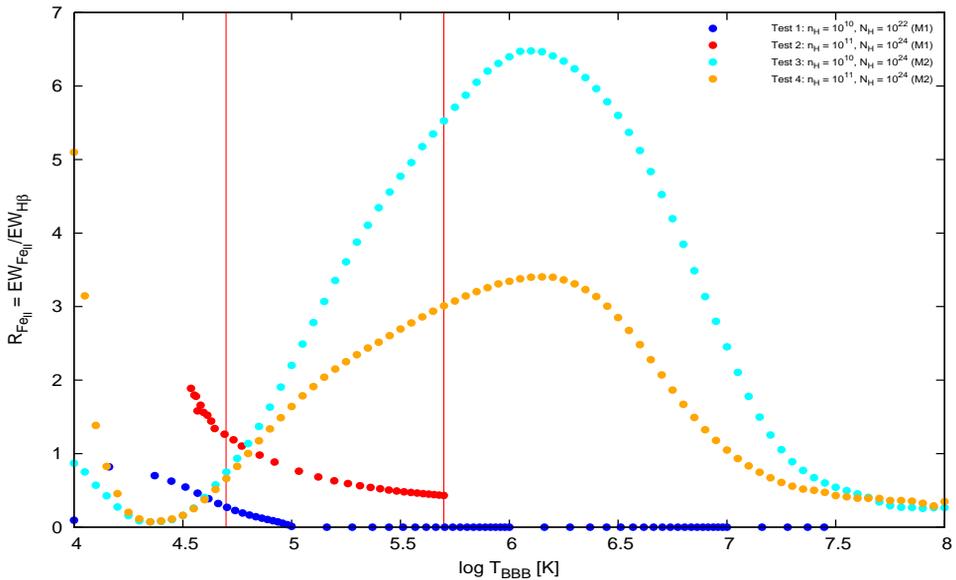}
\end{center}\caption{Comparison between  $\mathrm{R}_{\mathrm{Fe{II}}}$ - T$_{\mathrm{BBB}}$ for four different constant-density single cloud models. The blue ($\mathrm{n_H = 10^{10}\; cm^{-3}}$) and red ($\mathrm{n_H = 10^{11}\; cm^{-3}}$) dotted curves are for M1, the cyan ($\mathrm{n_H = 10^{10}\; cm^{-3}}$) and orange ($\mathrm{n_H = 10^{11}\; cm^{-3}}$) curves are for M2. Vertical red lines indicate the range of disk temperature considered [$5\times 10^4\;$K, $5\times 10^5\;$K] for which the black hole mass varies from $6.06\times 10^5\; \mathrm{M_{\odot}}$ to $6.06\times 10^9\; \mathrm{M_{\odot}}$.}\label{fig:R_Tbb_2}
\end{figure}

In M1, the dependence of $\mathrm{R_{FeII} - T_{BBB}}$ is monotonic (the red curve with points) for the case (ii), while we see a turnover for the trend in case (i) at $1.45\times 10^4\; K$ (the blue curve with points) although the monotonic trend reappears after this turnover and continues to the limit of the $\mathrm{T_{BBB}}$ considered for that case, i.e., $10^{7.5}\; \mathrm{K}$. This upper limit for the maximum of the disk temperature is solely considered for the modelling. In \cite{panda17}, we overplot $\mathrm{R_{FeII}}$ values obtained from two observational data on the modelled trends. We found that the trends did not justify the results from the observations. We intended to re-evaluate the trends by adopting the prescription in M2.
\par 
In M2, we test the same two cases of changing $\mathrm{n_H}$ from (iii) $10^{10} \;\mathrm{cm^{-3}}$ to (iv) $10^{11} \;\mathrm{cm^{-3}}$,  keeping the hydrogen column density at $\mathrm{N_H = 10^{24} \;cm^{-2}}$. For both these cases, we now clearly see the turnover in the trend between $\mathrm{R_{FeII} - T_{BBB}}$ close to $10^6\; \mathrm{K}$ which was absent in case (i) and its presence was speculative in case (ii). In the considered range of $\mathrm{T_{BBB}}$, we now have a proportional dependence of $\mathrm{R_{FeII}}$ on $\mathrm{T_{BBB}}$ (the vertical red lines depict the said range). For values of $\mathrm{T_{BBB}} \leq 2.24\times 10^4\; \mathrm{K}$, we again see a rising trend which we suspect is due to the variation in the value of $\mathrm{\alpha_{ox}}$ which is biased by the observationally-derived Eq.(4), where the authors \citep{lusso17} preferentially selected bluer candidates. For the considered $\mathrm{T_{BBB}}$ range the $\mathrm{R_{FeII}}$ lies within [0.75, 5.53] for $\mathrm{n_H}$ = $10^{10} \;\mathrm{cm^{-3}}$, and [0.66, 3.01] for $\mathrm{n_H}$ = $10^{10} \;\mathrm{cm^{-3}}$. In case (iv) we see that the FeII strength goes down by a factor 2 compared to case (iii), i.e., FeII emission is suppressed with rising mean density. The maximum values of $\mathrm{R_{FeII}}$ obtained in M2 is 6.48 at $\mathrm{T_{BBB}} = 1.26\times 10^6\; \mathrm{K}$ corresponding to $\mathrm{M_{BH}}$ as low as $1.5\times 10^4\; \mathrm{M_{\odot}}$. From a preliminary analysis of the \cite{shen11} SDSS DR7 quasar catalog, when the sample is z-corrected (i.e. $0.1 \leq \mathrm{z} \leq 0.9$) and errors in determining $\mathrm{FeII}$ and $\mathrm{H\beta}$ fluxes is kept within 20$\%$, we do have the maximum value of $\mathrm{R_{FeII}}$ at 6.56.
\bibliographystyle{ptapap}
\bibliography{ptapap}
\end{document}